\documentclass[12pt]{article}
\usepackage{graphicx}
\usepackage{amsfonts,amssymb}
\usepackage{amsmath}
\usepackage{citehack,cite}
\usepackage[colorlinks,breaklinks,allcolors=blue]{hyperref}

\newcommand{\no}{\nonumber}
\newcommand{\non}{\nonumber \\}
\newcommand{\be}{\begin{equation}}
\newcommand{\ee}{\end{equation}}
\newcommand{\bea}{\begin{eqnarray}}
\newcommand{\eea}{\end{eqnarray}}
\newcommand{\lp}{\left (}
\newcommand{\rp}{\right )}
\newcommand{\lb}{\left [}
\newcommand{\rb}{\right ]}

\newcommand{\sli}{\sum\limits}
\newcommand{\ve}[1]{{\bf #1}}
\newcommand{\vk}{\ve{k}}
\newcommand{\cB}{{\cal B}}
\newcommand{\rhok}{\rho_{\ve{k}}}
\newcommand{\rhomk}{\rho_{-\ve{k}}}

\begin{document}

\begin{center}
{\bf MICROSCOPIC DESCRIPTION OF THE LIQUID-GAS COEXISTENCE CURVE
FOR MORSE FLUIDS IN THE IMMEDIATE VICINITY OF THE CRITICAL POINT}
\end{center}

\begin{center}
{\sc I.V. Pylyuk, M.P. Kozlovskii, R.V. Romanik\footnote{e-mail: romanik@icmp.lviv.ua}}
\end{center}

\begin{center}
{\it Institute for Condensed Matter Physics
of the National Academy of Sciences of Ukraine,
1~Svientsitskii Str., 79011 Lviv, Ukraine}
\end{center}

\vspace{0.5cm}

{\small
The present work is aimed at investigating the behavior of Morse fluids
in the immediate vicinity of the critical point within the framework of
a cell model. This region is of both fundamental and practical importance,
yet presents analytical challenges due to the significant influence of
order parameter fluctuations. An analytical procedure is developed to
construct the upper part of the liquid-gas coexistence curve and
calculate its diameter, incorporating the non-Gaussian (quartic)
distribution of fluctuations. An explicit expression is derived for
the temperature-dependent analytical term appearing in the expression
for the rectilinear diameter. The numerical evaluation of the relevant
quantities is carried out using Morse potential parameters
representative of sodium. The coexistence curve is constructed both
with and without the inclusion of the analytical temperature-dependent
term in the calculation. A specific condition is identified under
which the agreement between the presented binodal branches and
Monte Carlo simulation data from other study, extrapolated to
the immediate vicinity of the critical point, is improved. It is shown
that better agreement is achieved when the analytical term is included
in the calculation of the liquid branch and omitted in the gas branch.
The proposed analytical approach may provide useful insight for
the theoretical study of critical phenomena in more
complex fluid systems.
}

\vspace{0.5cm}



\section{Introduction}
\label{sec1}

Our calculations within the framework of the grand canonical ensemble
are performed in the immediate vicinity of the critical point, which is a problematic area for theoretical and experimental studies.
The region in the vicinity of the critical point is of both fundamental
and applied interest, yet remains difficult to analyze due to
the significant role of fluctuation effects.
Calculations in this region are challenging to carry out
(see, for example, Refs. \cite{okumura_00,singh}, where computer
simulation data for the upper parts of the coexistence curves in
the temperature-density plane are lacking in the immediate vicinity
of the critical point). The high critical temperatures characteristic
of many metals still cannot be reached experimentally \cite{apf_11}.

This paper continues the previous studies \cite{kpd118,p120,pkdd123,dkrp124},
based on a cell fluid model with the Morse interaction potential.
An analytical method has been developed to describe the upper part of
the liquid-gas coexistence curve (i.e., the top of the coexistence dome)
for Morse fluids, taking into account the non-Gaussian (quartic)
distribution of order parameter fluctuations.
The order parameter and the coexistence curve diameter, both playing
an important role in the study of phase transitions, have been calculated.

The presence of a singularity in the coexistence curve diameter has been
theoretically predicted (see, for example, Refs. \cite{n181,jkh185,km109}).
However, the experimental observation of this effect is complicated by
the fact that the singular term may be masked by other contributions of
comparable magnitudes and similar exponents \cite{gdkg183}.
Since the exponent of the relative temperature in the singular term is
close to unity, the true singularity is difficult to distinguish
from the analytical contribution.
This anomaly in the diameter appears just below the critical point
and is challenging to capture in molecular simulations \cite{okumura_00}.
In their molecular dynamics study, the authors of Ref. \cite{okumura_00}
applied the equation corresponding to the law of the rectilinear diameter.
The behavior of coexistence curve diameters in fluids remains an active
topic of current research (see, for example,
Refs. \cite{glmbh118,bbd120,t124}).

In the present study, analytical expressions necessary for constructing
the top of the liquid-gas coexistence curve and for analyzing its diameter have been obtained. The numerical
values of the relevant quantities are provided.

\section{Basic expressions}
\label{sec2}

The theoretical description of the critical behavior of a simple fluid
is carried out within the framework of a cell model.
We consider an open system of classical interacting particles contained within a volume $V$. This volume is divided into $N_v$ cells, each of volume $v=V/N_v=c^3$, where $c$ denotes the linear size of a cell.
In contrast to the cell gas model described in \cite{rebenko_13,rebenko_15}, where each cell is assumed to contain at most one particle or to be empty, our approach allows a cell to contain multiple particles.

In terms of the collective variables $\rho_{\vk}$, the expression for
the grand partition function of the cell fluid model in the approximation
of the simplest non-Gaussian quartic fluctuation distribution
(the $\rho^4$ model) can be written as
\cite{kpd118,p120,kdp117}
\bea
&&
\Xi = g_W e^{N_v( E_\mu-a_0)} \int (d\rho)^{N_v} \exp\Biggl[ M N_v^{1/2} \rho_0
- \frac{1}{2} \sli_{\vk\in\cB} d(k) \rhok\rhomk \non
&&
- \frac{a_4}{4!} N_v^{-1} \sli_{{\vk_1,\ldots,\vk_4}\atop{\vk_i\in\cB}}
\rho_{\vk_1}\cdots\rho_{\vk_4} \delta_{\vk_1+\cdots+\vk_4}\Biggr].
\label{fpe_2d1fb}
\eea
Here
\bea
&&
g_W = \prod_{\vk\in\cB} \lp 2\pi \beta W(k)\rp^{-1/2}, \non
&&
E_\mu = - \frac{\beta W(0)}{2} (M + \tilde a_1)^2 + M a_{34} +
\frac{1}{2} d(0) a_{34}^2 - \frac{a_4}{24} a_{34}^4, \non
&&
a_{34} = - a_3 / a_4, \non
&&
M = \mu/W(0) - \tilde a_1, \quad \tilde a_1 = a_1 + d(0) a_{34} +
\frac{a_4}{6} a_{34}^3, \non
&&
d(k) = \frac{1}{\beta W(k)} - \tilde a_2, \quad
\tilde a_2 = \frac{a_4}{2} a_{34}^2 - a_2,
\label{fpe_2d2fb}
\eea
and $\delta_{\vk_1+\cdots+\vk_4}$ is the Kronecker symbol.
The wave vector $\vk$ belongs to the set
\[
\cB = \Big\{ \vk \!\!=\!\! (k_1, k_2, k_3)\Big| k_{i} \!\!=\!\! -\frac{\pi}{c}+
\frac{2\pi}{c}\frac{n_i}{N_a};
n_i=1,2,\ldots,N_a; i=1,2,3; N_v=N_a^3 \Big\},
\]
where $N_a$ is the number of cells along each axis.
The quantity $M$, linearly related to the thermodynamic chemical potential $\mu$, will hereafter be treated as the chemical potential.
Above, $\beta=1/(k_B T)$ is the inverse temperature, $k_B$ is the Boltzmann constant, and $T$ is the temperature. The Fourier transform of
the effective interaction potential $W(k)$ includes the Fourier transforms
of the attractive and repulsive parts of the Morse potential, which is
defined as
\be
\tilde U_{l_{12}} = \Psi_{l_{12}} - U_{l_{12}}; \quad
\Psi_{l_{12}} = D e^{-2(l_{12}- 1)/\alpha_R}, \quad
U_{l_{12}} = 2 D e^{-(l_{12}- 1)/\alpha_R}.
\label{fpe_2d3fb}
\ee
Here $\alpha_R = \alpha / R_0$, and $\alpha$ is  the effective interaction
radius. The parameter $R_0$ corresponds to the minimum of the function
$\tilde U_{l_{12}}$, and $D$ determines the depth of a potential
well. The Morse potential $\tilde U_{l_{12}}$ is a function of
the distance $l_{12}= |\ve{l}_{1}-\ve{l}_{2}|$ between cells $\ve{l}_{1}$
and $\ve{l}_{2}$. For convenience, all lengths are measured in $R_0$-units. As a result, $R_0$- and $R_0^3$-units
are used for the linear size of each cell $c$ and volume $v$, respectively.

The coefficients \cite{p120,kdp117}
\bea
&&
a_0 (T) = - \ln{T_0 (v,p(T))}, \quad
a_1 (T) = - \frac{T_1 (v,p(T))}{T_0 (v,p(T))}, \non
&&
a_2 (T) = - \frac{T_2 (v,p(T))}{T_0 (v,p(T))} + a_1^2(T),
\label{fpe_2d4fb} \\
&&
a_3 (T) = - \frac{T_3 (v,p(T)}{T_0 (v,p(T))}) - a_1^3(T) + 3 a_1(T) a_2(T), \non
&&
a_4 (T) = - \frac{T_4 (v,p(T))}{ T_0 (v,p(T))} + a_1^4(T) - 6 a_1^2(T) a_2(T)
+ 4 a_1(T) a_3(T) + 3 a_2^2(T) \no
\eea
appearing in the expression (\ref{fpe_2d1fb}) for $\Xi$ are
presented in terms of special functions
\be
T_n(v,p(T)) = \sli_{m=0}^\infty \frac{v^m}{m!} m^n e^{-p(T) m^2},
\label{fpe_2d5fb}
\ee
which are rapidly convergent series due to the condition $p(T)>0$.
The parameter
\be
p(T) = \frac{1}{2}\beta \chi \Psi (0),
\label{fpe_2d6fb}
\ee
is proportional to the Fourier transform
$\Psi(0)= D \pi \alpha_R^3 e^{2R_0/\alpha}/v$ of
the repulsive part of the interaction potential at
$k=0$ \cite{kpd118,p120,kdp117}. The positive parameter
	$\chi$ forms the Jacobian of transition from individual coordinates
	to collective variables. Thus,
the quantities $a_n (T)$ are functions of temperature and microscopic
parameters of the interaction potential, in particular,
of the ratio $R_0/\alpha$ characterizing real substances \cite{singh,lkg167}. Further, for convenience, we will
denote these quantities $a_n (T)$ as $a_n$.

Having the final formula for the logarithm of the grand partition
function \cite{kpd118,p120}, we can find the average number of particles
\be
\bar N = \frac{\partial \ln\Xi}{\partial \beta\mu}.
\label{fpe_2d7fb}
\ee
This relation allows us to express the chemical potential in terms
of the average number of particles $\bar N$ or in terms of
the average density
\be
\bar n = \frac{\bar N}{N_v} = \lp \frac{\bar N}{V}\rp v,
\label{fpe_2d8fb}
\ee
where $v$ is the volume of a cubic cell.

\section{Relationship between the density and the chemical potential
of fluid}
\label{sec3}

The nonlinear equation, which
describes the relationship between the density $\bar n$ and the chemical
potential $M$, can be represented as \cite{p120,pk722}
\be
\bar n = n_g - M + \sigma_{00}^{(-)} \lp \tilde h^2 + h^2_{cm}
\rp^{\frac{d-2}{2(d+2)}}.
\label{fpe_3d1fb}
\ee
Here $d = 3$ is the space dimension. The quantity
\be
\tilde h = M (\beta W(0))^{1/2}
\label{fpe_3d2fb}
\ee
is proportional to the chemical potential.
The quantity
\be
h_{cm} = \tilde\tau_1^{\ln E_1/\ln E_2}
\label{fpe_3d3fb}
\ee
is characterized by the renormalized relative temperature
\be
\tilde\tau_1 = -\tau \frac{c_{11}}{q} E_2^{n_0},
\label{fpe_3d4fb}
\ee
where $\tau = (T-T_c)/T_c$ ($T_c$ is the critical temperature),
and $E_l$ are eigenvalues of the renormalization
group linear transformation matrix. The quantity $c_{11}$
characterizes one of the coefficients in the solutions of recurrence
relations for the $\rho^4$ model, $n_0$ is the difference between
the points of the exit of the system from the critical fluctuation
regime at $T>T_c$ and $T<T_c$, and $q$ is associated with the averaging
of the wave vector square. Additional information about these parameters
can be found in Refs. \cite{kpd118,p120}.
The term
\be
n_g = - a_1 - a_2 a_{34} + \frac{a_4}{3} a_{34}^3
\label{fpe_3d5fb}
\ee
appearing in Eq. (\ref{fpe_3d1fb}) is determined by the coefficients $a_n$,
which are given in Eqs. (\ref{fpe_2d4fb}) and are included in the initial
expression (\ref{fpe_2d1fb}) for the grand partition function.
The coefficient $\sigma_{00}^{(-)}$ in Eq. (\ref{fpe_3d1fb}) is
a function of the quantity $\alpha_m = \tilde h/h_{cm}$,
which includes the chemical potential $M$ and the relative
temperature $\tau$ (see Ref. \cite{p120}).

In this paper, the expression (\ref{fpe_3d1fb}) for $\bar n$ at $M = -0$ and
$M = +0$ will be used to derive the equations describing the liquid-gas
coexistence curve in the temperature-density plane. Based on these
equations, we will obtain explicit analytical expressions for the order
parameter of the system and the diameter of the coexistence curve.
Let us start the calculations.

Using the relation $\beta = \beta_c (1 + \tau)^{-1}$ and singling
out temperature explicitly in Eq. (\ref{fpe_2d6fb}), we arrive at the
following expression accurate to within $\tau\ll 1$:
\be
p(T) = p_0 (1 - \tau).
\label{fpe_3d6fb}
\ee
Here $p_0 = \beta_c \chi \Psi (0)/2$. Substituting expression
(\ref{fpe_3d6fb}) for $p(T)$ into Eq. (\ref{fpe_2d5fb}), and retaining terms within the linear approximation in $\tau$, leads to
the following relation
\be
T_n(v,p(T)) = T_n(v,p_0) + T_{n+2}(v,p_0) p_0 \tau.
\label{fpe_3d7fb}
\ee
Taking into account Eq. (\ref{fpe_3d7fb}) and introducing the notations
\be
T_n(v,p_0) \equiv T_n, \quad \frac{T_m(v,p_0)}{T_n(v,p_0)} \equiv T_{mn},
\label{fpe_3d8fb}
\ee
we obtain the following expressions for the coefficients $a_n$
[see Eqs. (\ref{fpe_2d4fb})]:
\bea
&&
a_0 = a_{0c} + a_0^{(1)} \tau, \quad
a_1 = a_{1c} + a_1^{(1)} \tau, \quad
a_2 = a_{2c} + a_2^{(1)} \tau, \non
&&
a_3 = a_{3c} + a_3^{(1)} \tau, \quad
a_4 = a_{4c} + a_4^{(1)} \tau.
\label{fpe_3d9fb}
\eea
Here
\bea
&&
a_{0c} = -\ln T_0, \quad a_0^{(1)} = - T_{20} p_0, \non
&&
a_{1c} = - T_{10}, \quad a_1^{(1)} = a_{1c} (- T_{20} + T_{31}) p_0, \non
&&
a_{2c} = - T_{20} + a_{1c}^2, \quad
a_2^{(1)} = 2 a_{1c} a_1^{(1)} + (T_{20} - T_{42}) T_{20} p_0, \non
&&
a_{3c} = - T_{30} - a_{1c}^3 + 3 a_{1c} a_{2c},
\label{fpe_3d10fb} \\
&&
a_3^{(1)} = 3 (a_{1c}^2 a_1^{(1)} + a_{1c} a_2^{(1)}
+ a_1^{(1)} a_{2c}) + (T_{20} - T_{53}) T_{30} p_0, \non
&&
a_{4c} = - T_{40} + a_{1c}^4 - 6 a_{1c}^2 a_{2c}
+ 4 a_{1c} a_{3c} + 3 a_{2c}^2, \non
&&
a_4^{(1)} = 4 a_{1c}^3 a_1^{(1)} - 6 (a_{1c}^2 a_2^{(1)}
+ 2 a_{1c} a_1^{(1)} a_{2c}) + 4 (a_{1c} a_3^{(1)} + a_1^{(1)} a_{3c}) \non
&&
+ 6 a_{2c} a_2^{(1)} + (T_{20} - T_{64}) T_{40} p_0. \no
\eea
According to Eq. (\ref{fpe_3d5fb}), the explicit dependence of
the quantity $n_g$ on the relative temperature $\tau$ can be obtained
using Eqs. (\ref{fpe_3d9fb}) as well as the relation
\be
a_{34} = a_{34c} + a_{34}^{(1)} \tau,
\label{fpe_3d11fb}
\ee
where
\be
a_{34c} = - \frac{a_{3c}}{a_{4c}}, \quad
a_{34}^{(1)} = a_{34c} \lp \frac{a_3^{(1)}}{a_{3c}} -
\frac{a_4^{(1)}}{a_{4c}}\rp.
\label{fpe_3d12fb}
\ee
We arrive at the expression
\be
n_g = n_{gc} + n_g^{(1)} \tau.
\label{fpe_3d13fb}
\ee
Here
\bea
&&
n_{gc} = - a_{1c}- a_{2c} a_{34c} + \frac{a_{4c}}{3} a_{34c}^3,
\label{fpe_3d14fb} \\
&&
n_g^{(1)} = - a_1^{(1)} - a_{2c} a_{34}^{(1)} - a_2^{(1)} a_{34c} +
\frac{1}{3} \lp 3 a_{4c} a_{34c}^2 a_{34}^{(1)} + a_{34c}^3 a_4^{(1)}\rp. \no
\eea
The components of the quantities $a_n$, $a_{34}$, and $n_g$ are given in
Tables~\ref{tab_1fb_fpe}, \ref{tab_2fb_fpe}, and \ref{tab_3fb_fpe} for
the set of parameters $R_0/\alpha = 2.9544$ (which is typical of
sodium (Na), see Refs. \cite{singh,lkg167}) and $\chi = 1.1243$,
$p_0 = 1.8100$, $v = 2.4191$ (see Refs. \cite{kpd118,p120,kdp117}).
\begin{table}[htbp]
\caption{Values of the quantities determining the coefficients
$a_0$, $a_1$, and $a_2$.}
\label{tab_1fb_fpe}
\begin{center}
\begin{tabular}{cccccc}
\hline
\multicolumn{1}{c}{$a_{0c}$} & \multicolumn{1}{c}{$a_0^{(1)}$} &
\multicolumn{1}{c}{$a_{1c}$} & \multicolumn{1}{c}{$a_1^{(1)}$} &
\multicolumn{1}{c}{$a_{2c}$} & \multicolumn{1}{c}{$a_2^{(1)}$} \\
\hline
$-0.3350$ & $-0.5234$ & $-0.2862$ & $-0.3845$ & $-0.2073$ & $-0.1846$ \\
\hline
\end{tabular}
\end{center}
\end{table}
\begin{table}[htbp]
\caption{Numerical estimates of the quantities determining the coefficients
$a_3$, $a_4$, and $a_{34}$.}
\label{tab_2fb_fpe}
\begin{center}
\begin{tabular}{cccccc}
\hline
\multicolumn{1}{c}{$a_{3c}$} & \multicolumn{1}{c}{$a_3^{(1)}$} &
\multicolumn{1}{c}{$a_{4c}$} & \multicolumn{1}{c}{$a_4^{(1)}$} &
\multicolumn{1}{c}{$a_{34c}$} & \multicolumn{1}{c}{$a_{34}^{(1)}$} \\
\hline
$-0.0938$ & $-0.1419$ & 0.0376 & 0.4110 & 2.4925 & $-23.4516$ \\
\hline
\end{tabular}
\end{center}
\end{table}
\begin{table}[htbp]
\caption{Numerical values of the quantities $n_{gc}$, $n_g^{(1)}$,
as well as the roots $\sigma_{030}^{'(-)}$, $\sigma_{010}^{'(+)}$
of a specific cubic equation and the coefficients
$\sigma_{00c}^{(-)}(\sigma_{030}^{'(-)})$,
$\sigma_{00c}^{(-)}(\sigma_{010}^{'(+)})$
in Eqs. (\ref{fpe_5d1fb}), (\ref{fpe_5d2fb}).}
\label{tab_3fb_fpe}
\begin{center}
\begin{tabular}{cccccc}
\hline
\multicolumn{1}{c}{$n_{gc}$} & \multicolumn{1}{c}{$n_g^{(1)}$} &
\multicolumn{1}{c}{$\sigma_{030}^{'(-)}$} &
\multicolumn{1}{c}{$\sigma_{010}^{'(+)}$} &
\multicolumn{1}{c}{$\sigma_{00c}^{(-)}(\sigma_{030}^{'(-)})$} &
\multicolumn{1}{c}{$\sigma_{00c}^{(-)}(\sigma_{010}^{'(+)})$} \\
\hline
0.9971 & $-7.3776$ & $-3.3745$ & 3.3745 & $-0.8359$ & 0.8359 \\
\hline
\end{tabular}
\end{center}
\end{table}
As is seen from Table~\ref{tab_3fb_fpe}, the quantities
$\sigma_{030}^{'(-)}$ and $\sigma_{010}^{'(+)}$, as well as
$\sigma_{00c}^{(-)}(\sigma_{030}^{'(-)})$ and
$\sigma_{00c}^{(-)}(\sigma_{010}^{'(+)})$,  have equal magnitudes
but opposite signs. This property will be used in Section~\ref{sec5}
in the manipulations involving Eqs. (\ref{fpe_5d1fb})
and (\ref{fpe_5d2fb}).

Let us now consider several limiting cases and write the expressions for
the average density $\bar n$ for them.

\section{Density for some limiting cases}
\label{sec4}

The equation (\ref{fpe_3d1fb}), taking into account the expression
(\ref{fpe_3d13fb}), takes the form
\be
\bar n = n_{gc} + n_g^{(1)} \tau - M + \sigma_{00}^{(-)}
\lp \tilde h^2 + h^2_{cm}\rp^{\frac{d-2}{2(d+2)}}.
\label{fpe_4d1fb}
\ee
The general form of Eq. (\ref{fpe_4d1fb}) naturally allows for
the transition to cases where one of the variables (temperature or
chemical potential) is the determining factor in describing the behavior
of the order parameter.

Let us describe the behavior of $\bar n$ for some limiting cases.
One such case is the absence of chemical potential $M$ ($M = 0$, and hence
$\tilde h = 0$) while $T\neq T_c$. Then we have
\be
\sigma_{00}^{(-)}\Big|_{M=0} = \lp 1 + \frac{1}{2} \tau\rp
\sigma_{00c}^{(-)}(\sigma'_0),
\label{fpe_4d2fb}
\ee
where
\be
\sigma_{00c}^{(-)}(\sigma'_0) = \frac{1}{(\beta_c W(0))^{1/2}}
\lb e_0^{(-)} + \frac{f_{Iv}}{s^3}\rb,
\label{fpe_4d3fb}
\ee
and $e_0^{(-)} = \sigma'_0 s^{-1/2}$. The quantity $\sigma'_0$ is
a solution of a specific cubic equation (see Refs. \cite{p120,pk722}).
The form of the solutions of this equation depends on
the sign of the discriminant. The renormalization group parameter $s$
determines the division of the phase space of collective variables into
layers \cite{ymo287,YuKP_2001}. The expression for $f_{Iv}$ is given
in \cite{p120}. Based on Eq. (\ref{fpe_4d1fb}), we obtain the dependence
\be
\bar n = n_{gc} + n_g^{(1)} \tau + \sigma_{00}^{(-)}\Big|_{M=0}
\tilde\tau_1^{\beta}.
\label{fpe_4d4fb}
\ee
Here the critical exponent $\beta=\nu/2$ is determined by the critical
exponent of the correlation length $\nu$ \footnote{It is conventional in physics to denote the inverse temperature by $\beta$, while the same symbol is also commonly used for the order parameter critical exponent. The intended meaning will be evident from the context.}. The quantity $\tilde\tau_1$
is defined in Eq. (\ref{fpe_3d4fb}).

Another limiting case is $M\neq 0$ and $T = T_c$ ($\tau = 0$,
and hence $\tilde\tau_1 = 0$). According to Eq. (\ref{fpe_4d1fb}),
the density $\bar n$ at $T=T_c$ satisfies the equality
\be
\bar n = n_{gc} - M + \sigma_{00}^{(-)}\Big|_{T=T_c} \tilde h^{1/\delta},
\label{fpe_4d5fb}
\ee
where
\be
\sigma_{00}^{(-)}\Big|_{T=T_c} = \frac{6}{5} \frac{1}{(\beta_c W(0))^{1/2}}
\lb e_0^{(-)} + \gamma_s^{(-)} - e_2^{(-)} \rb,
\label{fpe_4d6fb}
\ee
and $\tilde h = M (\beta_c W(0))^{1/2}$, $\delta = (d + 2)/(d - 2)$.
The quantity $e_2^{(-)}$, like $e_0^{(-)}$, depends on $\sigma'_0$.
The coefficient $\gamma_s^{(-)}$ characterizes
the non-analytical contribution to the thermodynamic potential.
The expressions for these quantities are given in Ref. \cite{p120}.

In the general case ($M\neq 0$ and $T\neq T_c$),
Eq. (\ref{fpe_4d1fb}) can be rewritten as
\be
\bar n = n_{gc} + n_g^{(1)} \tau - M + \sigma_{00}^{(-)}
\lp \tilde h^2 + \tilde\tau_1^{2\beta\delta} \rp^{1/(2\delta)}.
\label{fpe_4d7fb}
\ee
Note that $M\ll 1$, and $\tilde h\sim M$. Therefore, the term $M$
in the right-hand sides of Eqs. (\ref{fpe_4d1fb}), (\ref{fpe_4d5fb}),
and (\ref{fpe_4d7fb}) is significantly smaller than the last term
and can be neglected.

\section{Equation for the order parameter and the diameter of
the coexistence curve}
\label{sec5}

Let us proceed to deriving the equations that describe the binodal curve
in a narrow temperature interval near the critical point. Numerical calculations will be
illustrated by an example of the parameters of the Morse interaction
potential, which are characteristic of sodium (Na) and are given
in Section~\ref{sec3}. Based on Eqs. (\ref{fpe_3d4fb}),
(\ref{fpe_4d2fb}), and (\ref{fpe_4d4fb}), we obtain the following
formulas for the gas density $\bar n_G$ and
the liquid density $\bar n_L$:
\be
\bar n_G = n_{gc} - n_g^{(1)} (-\tau) +
\lp \frac{c_{11}}{q} E_2^{n_0}\rp^{\beta}
\sigma_{00c}^{(-)}(\sigma_{030}^{'(-)}) (-\tau)^{\beta}
\label{fpe_5d1fb}
\ee
and
\be
\bar n_L = n_{gc} - n_g^{(1)} (-\tau) +
\lp \frac{c_{11}}{q} E_2^{n_0}\rp^{\beta}
\sigma_{00c}^{(-)}(\sigma_{010}^{'(+)}) (-\tau)^{\beta}.
\label{fpe_5d2fb}
\ee
Note that $\tau< 0$ and $|\tau|\ll 1$.
In the right-hand sides of Eqs. (\ref{fpe_5d1fb}) and (\ref{fpe_5d2fb}),
we have neglected the term proportional to $(-\tau)^{\beta + 1}$ since
it is less significant than the term proportional to $(-\tau)^{\beta}$.
The roots $\sigma_{030}^{'(-)}$ and $\sigma_{010}^{'(+)}$ of a specific
cubic equation (see Ref. \cite{pk722} and Table~\ref{tab_3fb_fpe}) refer
to the chemical potentials $M = -0$ and $M = +0$, respectively.
The coefficients $\sigma_{00c}^{(-)}(\sigma_{030}^{'(-)})$ and
$\sigma_{00c}^{(-)}(\sigma_{010}^{'(+)})$ can be obtained from
Eq. (\ref{fpe_4d3fb}) by substituting the quantities
$\sigma_{030}^{'(-)}$ and $\sigma_{010}^{'(+)}$, respectively,
in place of $\sigma'_0$. For convenience, numerical values of the other coefficients appearing in Eqs.~(\ref{fpe_5d1fb}) and~(\ref{fpe_5d2fb}) are given in Table~\ref{tab_3afb_fpe}.

\begin{table}[htbp]
	\caption{The quantities $c_{11}$, $q$, $E_2$, $n_0$, and
		$\beta$ appearing in Eqs. (\ref{fpe_5d1fb}) and (\ref{fpe_5d2fb}).}
	\label{tab_3afb_fpe}
	\begin{center}
		\begin{tabular}{ccccc}
			\hline
			\multicolumn{1}{c}{$c_{11}$} & \multicolumn{1}{c}{$q$} &
			\multicolumn{1}{c}{$E_2$} & \multicolumn{1}{c}{$n_0$} &
			\multicolumn{1}{c}{$\beta$} \\
			\hline
			0.9425 & 1.2356 & 8.3079 & 0.5000 & 0.3024 \\
			\hline
		\end{tabular}
	\end{center}
\end{table}

Considering that $\sigma_{00c}^{(-)}(\sigma_{030}^{'(-)}) =
- \sigma_{00c}^{(-)}(\sigma_{010}^{'(+)})$ (see Table~\ref{tab_3fb_fpe})
and using Eqs. (\ref{fpe_5d1fb}) and (\ref{fpe_5d2fb}), we arrive at
the scaling equation
\be
\frac{1}{2} (\bar n_L - \bar n_G) = B_0 |\tau|^{\beta}.
\label{fpe_5d3fb}
\ee
Here
\be
B_0 = \lp \frac{c_{11}}{q} E_2^{n_0}\rp^{\beta}
\sigma_{00c}^{(-)}(\sigma_{010}^{'(+)}).
\label{fpe_5d4fb}
\ee
The order parameter of the studied continuous system is defined as
\be
\frac{\bar n_L - \bar n_G}{n_{gc}} = \frac{2 B_0}{n_{gc}} |\tau|^{\beta}.
\label{fpe_5d5fb}
\ee
The equation describing the law of rectilinear diameter of
the liquid-gas coexistence curve has the form
\be
\frac{\bar n_L + \bar n_G}{2} = n_{gc} + D_1 |\tau|.
\label{fpe_5d6fb}
\ee
For the coefficient of the term linear in $\tau$, we find
\be
D_1 = - n_g^{(1)}.
\label{fpe_5d7fb}
\ee
The expression for $n_g^{(1)}$, as well as for $n_{gc}$, is presented in
Section~\ref{sec3} [see Eqs. (\ref{fpe_3d14fb})]. The quantities
$B_0$ and $D_1$ are given in Table~\ref{tab_4fb_fpe}.

\begin{table}[htbp]
\caption{The coefficient $B_0$ in expression (\ref{fpe_5d5fb}) for
the order parameter and the coefficient $D_1$ in equation (\ref{fpe_5d6fb})
for the rectilinear diameter.}
\label{tab_4fb_fpe}
\begin{center}
\begin{tabular}{cc}
\hline
\multicolumn{1}{c}{$B_0$} & \multicolumn{1}{c}{$D_1$} \\
\hline
1.0608 & 7.3776 \\
\hline
\end{tabular}
\end{center}
\end{table}

\begin{figure}[htbp]
	\centering \includegraphics[width=0.65\textwidth]{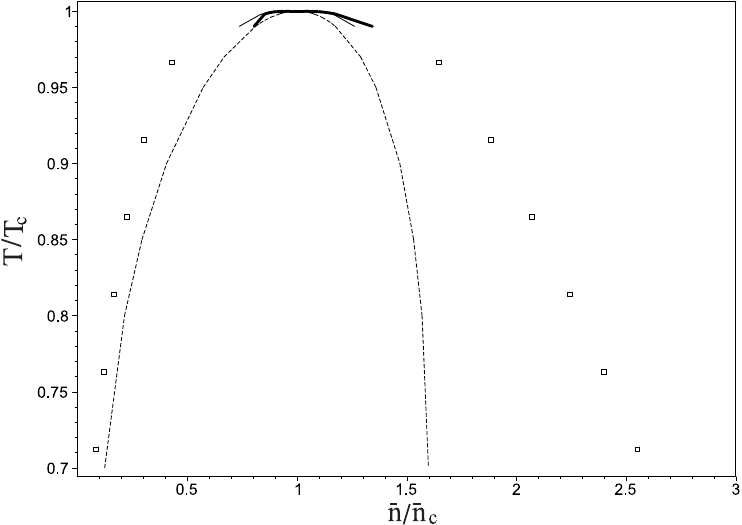}
	\caption{Upper part of the binodal for sodium in
		the temperature-density plane. The bold solid line represents the result of a calculation that includes an analytical temperature-dependent term.
		The thin solid line shows the result obtained without this term.
		The dashed line represents the zero-mode approximation \cite{kdp117}.
		Boxes indicate Monte Carlo simulation data \cite{singh}.}
	\label{fig_1fb_fpe}
\end{figure}

It should be noted that the present approach is aimed at calculating and
analyzing the characteristics of the system in a narrow vicinity of $T_c$
($|\tau|<\tau^*\sim 10^{-2}$, see \cite{ymo287,YuKP_2001}), where
theoretical and experimental research is difficult to carry out.

The upper part of the liquid-gas coexistence curve, obtained from
Eqs. (\ref{fpe_5d1fb}) and (\ref{fpe_5d2fb}) for the parameters
of the Morse interaction potential corresponding to sodium (Na),
is shown in Fig.~\ref{fig_1fb_fpe} as a bold solid line. This line
corresponds to the case where the calculation of the relevant quantities, particularly the coefficients $a_n(T)$ from Eq.~(\ref{fpe_3d9fb}), includes the contribution from the analytical term proportional to $\tau$ (i.e., the linear temperature dependence). The thin solid
line in Fig.~\ref{fig_1fb_fpe} corresponds to the case where
the analytical term is neglected in the calculation (in particular,
the coefficients $a_n(T)$ are taken at $T=T_c$).
For comparison, the coexistence curve obtained for Na within
the so-called zero-mode approximation \cite{kdp117}, which corresponds
to the mean-field approximation, is also shown in the figure as a dashed
line. The data obtained by Monte Carlo simulations \cite{singh} are
shown as boxes.

Figure~\ref{fig_1fb_fpe} shows which curve better agrees with the extrapolated data for Na, based on computer simulations \cite{singh} extended to $T/T_c \approx 1$. The zero-mode approximation (dashed line) does not
account for order parameter fluctuations and is ineffective in
the immediate vicinity of $T_c$. For the liquid branch of the binodal,
better agreement is provided by the bold solid line (the case with
analytical temperature-dependent term included). The gas branch of
the binodal is better approximated (i.e., closer to the Monte Carlo
results) by the thin solid line (the case where analytical
temperature-dependent term is neglected).

\begin{figure}[htbp]
	\centering \includegraphics[width=0.65\textwidth]{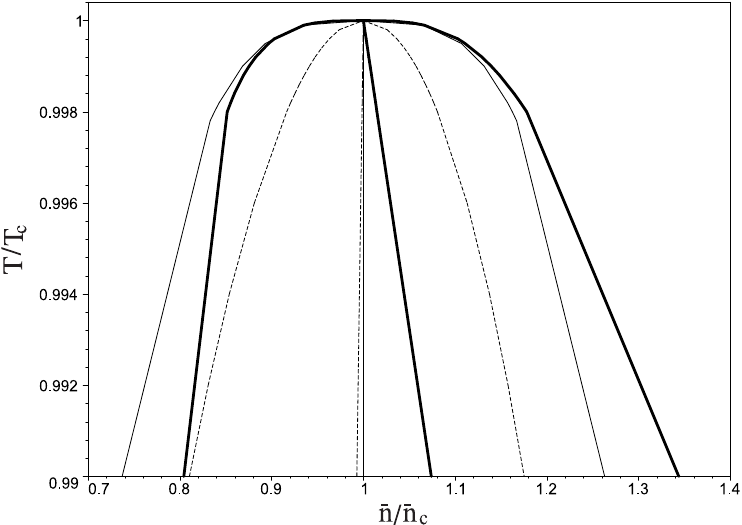}
	\caption{Liquid-gas coexistence curves and their diameters obtained
		in the immediate vicinity of the critical point using different
		calculation methods. Curve notations are the same as
		in Fig.~\ref{fig_1fb_fpe}.}
	\label{fig_2fb_fpe}
\end{figure}

\begin{figure}[htbp]
	\centering \includegraphics[width=0.65\textwidth]{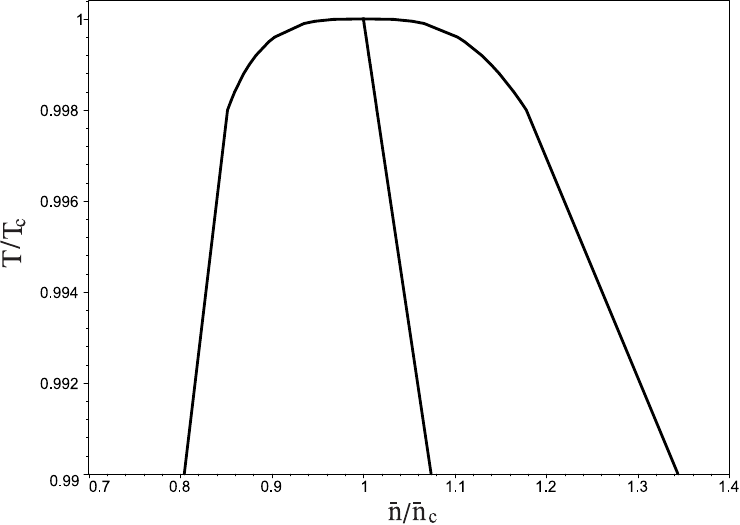}
	\caption{Upper part of the liquid-gas coexistence curve and its diameter,
		obtained with analytical temperature-dependent term included
		in both the liquid and gas branch calculations.}
	\label{fig_3fb_fpe}
\end{figure}
\begin{figure}[htbp]
	\centering \includegraphics[width=0.65\textwidth]{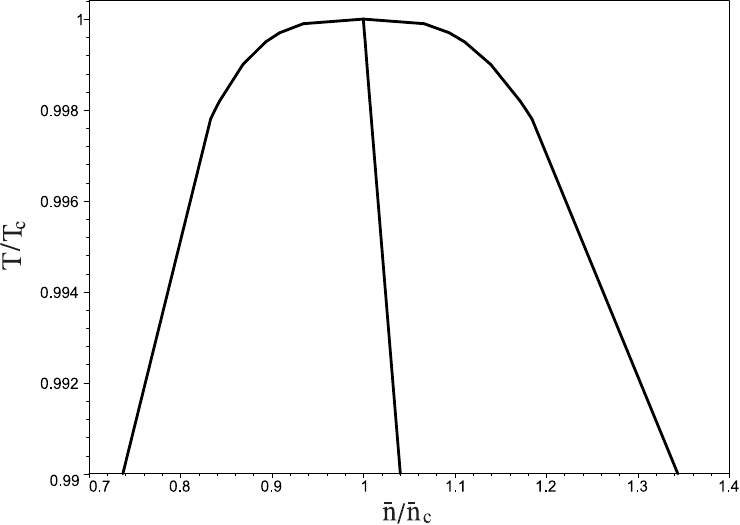}
	\caption{Upper part of the liquid-gas coexistence curve and its diameter,
		obtained with analytical temperature-dependent term included
		in the liquid branch calculation but omitted in the gas branch.}
	\label{fig_4fb_fpe}
\end{figure}

The binodal curves from Fig.~\ref{fig_1fb_fpe}, along with their
corresponding diameters, are shown in Fig.~\ref{fig_2fb_fpe} in
the immediate vicinity of the critical point. A finer scale is
used along the ordinate axis compared to Fig.~\ref{fig_1fb_fpe}.

The upper parts of the binodals along with their corresponding
diameters are shown separately in Figs.~\ref{fig_3fb_fpe} and
\ref{fig_4fb_fpe}, for different methods of computing the gas branch
of the binodal (see the respective figure captions for details).
As noted above, the gas branch of the binodal in Fig.~\ref{fig_4fb_fpe},
compared to that in Fig.~\ref{fig_3fb_fpe}, provides a better
reproduction of the Monte Carlo simulation results \cite{singh}
extrapolated to the immediate vicinity of the critical point.

\section{Conclusions}
\label{sec6}

The present work has focused on the investigation of Morse fluid behavior
near the critical point within the framework of a cell model.
The vicinity of the critical point is of both fundamental and practical
interest, while also being challenging to analyze due to the significant
role of fluctuation effects.

An analytical procedure has been developed for constructing the liquid-gas
coexistence curve and calculating its diameter in the critical
region. The numerical evaluation of the relevant quantities has been
illustrated using Morse potential parameters typical for sodium.
The critical point parameters for liquid alkali metals, sodium and
potassium, previously obtained within our approach
(see \cite{kpd118,p120,kdp117}), are consistent with
available experimental data.

The analysis of the relationship between density and chemical potential
at subcritical temperatures has enabled the derivation of equations
describing the liquid-gas coexistence curve in the temperature-density
plane. An explicit expression has been derived for the analytical
temperature-dependent term appearing in the equation for the rectilinear
diameter of the binodal. The upper part of the liquid-gas coexistence
curve and its diameter have been constructed both with and without
including the analytical temperature-dependent term in the calculation.
A condition has been established that improves the agreement of the binodal
branches with Monte Carlo simulation data \cite{singh} extrapolated to
the immediate vicinity of the critical point. It has been clearly
demonstrated that better agreement is achieved when the analytical
temperature-dependent term is neglected in the calculation of
the gas branch, but retained in the calculation of the liquid branch.

The analytical approach developed for a simple fluid system may prove
useful for investigating the critical behavior of multicomponent fluids.
The present study also represents a certain methodological contribution
to the theoretical description of critical phenomena.


\begin{thebibliography}{99}

\bibitem{okumura_00}
H. Okumura, F. Yonezawa, Liquid-vapor coexistence curves of several
interatomic model potentials, J. Chem. Phys. 113 (2000) 9162-9168,
\url{https://doi.org/10.1063/1.1320828}.

\bibitem{singh}
J.K. Singh, J. Adhikari, S.K. Kwak, Vapor--liquid phase
coexistence curves for Morse fluids, Fluid Phase Equilib. 248 (2006) 1-6,
\url{https://doi.org/10.1016/j.fluid.2006.07.010}.

\bibitem{apf_11}
E.M. Apfelbaum, The calculation of vapor-liquid coexistence curve of
Morse fluid: Application to iron, J. Chem. Phys. 134 (2011) 194506,
\url{https://doi.org/10.1063/1.3590201}.

\bibitem{kpd118}
M.P. Kozlovskii, I.V. Pylyuk, O.A. Dobush, The equation of state of
a cell fluid model in the supercritical region, Condens. Matter Phys.
21 (2018) 43502,
\url{https://doi.org/10.5488/CMP.21.43502}.

\bibitem{p120}
I.V. Pylyuk, Fluid critical behavior at liquid--gas
phase transition: Analytic method for microscopic description,
J. Mol. Liq. 310 (2020) 112933,
\url{https://doi.org/10.1016/j.molliq.2020.112933}.

\bibitem{pkdd123}
I.V. Pylyuk, M.P. Kozlovskii, O.A. Dobush, M.V. Dufanets,
Morse fluids in the immediate vicinity of the critical point: Calculation
of thermodynamic coefficients, J. Mol. Liq. 385 (2023) 122322,
\url{https://doi.org/10.1016/j.molliq.2023.122322}.

\bibitem{dkrp124}
O.A. Dobush, M.P. Kozlovskii, R.V. Romanik, I.V. Pylyuk, Thermodynamic
response functions in a cell fluid model, Ukr. J. Phys. 69 (2024) 919-929,
\url{https://doi.org/10.15407/ujpe69.12.919}.

\bibitem{n181}
J.F. Nicoll, Critical phenomena of fluids: Asymmetric
Landau-Ginz\-burg-Wilson model, Phys. Rev. A 24 (1981) 2203-2220,
\url{https://doi.org/10.1103/PhysRevA.24.2203}.

\bibitem{jkh185}
S. J\"ungst, B. Knuth, F. Hensel, Observation of singular diameters
in the coexistence curves of metals, Phys. Rev. Lett. 55 (1985) 2160-2163,
\url{https://doi.org/10.1103/PhysRevLett.55.2160}.

\bibitem{km109}
V.L. Kulinskii, N.P. Malomuzh, The nature of the rectilinear diameter
singularity, Physica A 388 (2009) 621-627,
\url{https://doi.org/10.1016/j.physa.2008.11.014}.

\bibitem{gdkg183}
S.C. Greer, B.K. Das, A. Kumar, E.S.R. Gopal, Critical behavior of
the diameters of liquid-liquid coexistence curves, J. Chem. Phys.
79 (1983) 4545-4552,
\url{https://doi.org/10.1063/1.446369}.

\bibitem{glmbh118}
Y. Garrabos, C. Lecoutre, S. Marre, D. Beysens, I. Hahn, Liquid-vapor
rectilinear diameter revisited, Phys. Rev. E 97 (2018) 020101,
\url{https://doi.org/10.1103/PhysRevE.97.020101}.

\bibitem{bbd120}
O. Bakai, M. Bratchenko, S. Dyuldya, On the singularity of the liquid-gas
coexistence curve diameter, Ukr. J. Phys. 65 (2020) 802-809,
\url{https://doi.org/10.15407/ujpe65.9.802}.

\bibitem{t124}
A.V. Tatarenko, Coexistence curve diameter and slope of vapor pressure:
``universal'' relation in the critical region, J. Mol. Liq.
394 (2024) 123776,
\url{https://doi.org/10.1016/j.molliq.2023.123776}.

\bibitem{rebenko_13}
A.L. Rebenko, ~Cell ~gas ~model ~of ~classical ~statistical ~systems,
Rev. Math. Phys. 25 (2013) 1330006,
\url{https://doi.org/10.1142/S0129055X13300069}.

\bibitem{rebenko_15}
V.A. Boluh, A.L. Rebenko. Cell gas free energy as an approximation of
the continuous model, J. Mod. Phys. 6 (2015) 168-175,
\url{https://doi.org/10.4236/jmp.2015.62022}.

\bibitem{kdp117}
M.P. Kozlovskii, O.A. Dobush, I.V. Pylyuk,
Using a cell fluid model for the description of a phase transition in
simple liquid alkali metals, Ukr. J. Phys. 62 (2017) 865-873,
\url{https://doi.org/10.15407/ujpe62.10.0865}.

\bibitem{lkg167}
R.C. Lincoln, K.M. Koliwad, P.B. Ghate, Morse-potential evaluation of
second- and third-order elastic constants of some cubic metals,
Phys. Rev. 157 (1967) 463-466,
\url{https://doi.org/10.1103/PhysRev.157.463}.

\bibitem{pk722}
I.V. Pylyuk, M.P. Kozlovskii, First-order phase
transition in the framework of the cell fluid model: Regions of
chemical potential variation and the corresponding densities,
Ukr. J. Phys. 67 (2022) 54-61,
\url{https://doi.org/10.15407/ujpe67.1.54}.

\bibitem{ymo287}
I.R. Yukhnovskii, Phase Transitions of the Second Order. Collective
Variables Method, World Scientific, Singapore, 1987.

\bibitem{YuKP_2001}
I.R. Yukhnovskii, M.P. Kozlovskii, I.V. Pylyuk,
Microscopic Theory of Phase Transitions in the Three-Dimensional Systems,
Eurosvit, Lviv, 2001 [in Ukrainian].
\end{thebibliography}
\end{document}